\documentclass[%
 reprint,
 amsmath,amssymb,
 aps,
 prb,
]{revtex4-1}

\usepackage{graphicx}
\usepackage{dcolumn}
\usepackage{bm}
\usepackage{natbib}

\usepackage{url}
\usepackage{xspace}
\usepackage{booktabs}
\usepackage{multirow}
\renewcommand{\,}{\hspace{1.5mm}}

\def\bal#1\eal{\begin{align*}#1\end{align*}}
\def\baln#1\ealn{\begin{align}#1\end{align}}

\renewcommand{\vec}[1]{\mathbf{#1}}

\newcommand{\ve}{\varepsilon}

\begin{document}


\title{Spontaneous breaking of time-reversal symmetry at the edges of 1T' monolayer transition metal dichalcogenides.}

\author{Line Jelver}
 \affiliation{Synopsys Denmark, Fruebjergvej 3, PostBox 4, DK-2100 Copenhagen, Denmark.}
 \altaffiliation[Line Jelver also at ]{CAMD, Dept. of Physics, Technical University of Denmark, Bldg. 307, DK-2800 Kongens Lyngby, Denmark.}

\author{Thomas Olsen}%
\affiliation{CAMD, Dept. of Physics, Technical University of Denmark, Bldg. 307, DK-2800 Kongens Lyngby, Denmark.}

\author{Daniele Stradi}%
  \affiliation{Synopsys Denmark, Fruebjergvej 3, PostBox 4, DK-2100 Copenhagen, Denmark.}

\author{Kurt Stokbro}%
  \affiliation{Synopsys Denmark, Fruebjergvej 3, PostBox 4, DK-2100 Copenhagen, Denmark.}

\author{Karsten Wedel Jacobsen}%
  \email{kwj@fysik.dtu.dk}
  \affiliation{CAMD, Dept. of Physics, Technical University of Denmark, Bldg. 307, DK-2800 Kongens Lyngby, Denmark.}


\date{\today}

\begin{abstract}
\section*{Abstract}

Using density functional theory calculations and the Greens's function formalism, we report the existence of magnetic edge states with a non-collinear spin texture present on different edges of the 1T’ phase of the three monolayer transition metal dichalcogenides (TMDs): MoS$_2$, MoTe$_2$ and WTe$_2$. The magnetic states are gapless and accompanied by a spontaneous breaking of the time-reversal symmetry. This may have an impact on the prospects of utilizing WTe$_2$ as a quantum spin Hall insulator. It has previously been suggested that the topologically protected edge states of the 1T' TMDs could be switched off by applying a perpendicular electric field\cite{Qian2014}. We confirm with fully self-consistent DFT calculations, that the topological edge states can be switched off. The investigated magnetic edge states are seen to be robust and remains gapless when applying a field.

\end{abstract}

\maketitle


\section{\label{sec:intro}Introduction}

In 2005, graphene was predicted to be a quantum spin Hall insulator\cite{Kane2005a} with a band gap emerging at the Dirac point as a consequence of spin-orbit coupling (SOC). However, due to the tiny spin-orbit coupling in graphene, it has not yet been possible to verify the prediction experimentally. The quantum spin Hall effect was, however, predicted and observed in HgTe quantum wells in 2007\cite{Mosfets2013,Konig2007} and the three-dimensional analogue of the effect was realized in Bi and Sb chalcogenides in 2009\cite{Hsieh2009,Zhang2009}. Meanwhile, several predictions for 2D materials exhibiting the quantum spin Hall effect had appeared. For example, the graphene-like materials silicene,\cite{Ezawa2012, Ezawa2013} germanene\cite{Zhang2016a, Amlaki2016} and stanene\cite{Tang2014} as well as certain monolayers of transition metal dichalcogenides (TMDs) in either the 1T' phase\cite{Qian2014} or the Haeckelite crystal structure\cite{Nie2015}. In contrast to other commonly occurring transition metal dichalcogenides, WTe$_2$ is of particular interest, since the 1T' structure is the naturally occurring phase, and monolayers can thus be obtained by direct exfoliation. Indeed, the quantum spin Hall effect in WTe$_2$ has recently been verified experimentally at temperatures up to 100 K\cite{Fei2017, Tang2017, Wu2018} and the band gap was reported to be about 50 meV.

Two-dimensional (2D) insulators with an electronic structure which is invariant under time-reversal symmetry can be characterized as being either trivial insulators or quantum spin Hall insulators\cite{Kane2005} (QSHIs). This distinction constitutes a topological $\mathbb{Z}_2$ classification, $\mathbb{Z}_2=0$ denoting the topologically trivial state and $\mathbb{Z}_2=1$ denoting the QSHI state. It is only possible for a material to change from one of these topological states to the other if either the gap closes or time-reversal symmetry is broken. For bulk materials, the topological state does not have any direct observational consequences. However, assuming that time-reversal symmetry is conserved, any interface between a trivial insulator and a QSHI e.i. between the two different topological states, is guaranteed to host gapless boundary states. Since vacuum can be regarded as a trivial insulator it follows that also any edge of a quantum spin Hall insulator hosts gapless states.

The gapless boundary or edge states in quantum spin Hall insulators are protected from back scattering by time-reversal symmetry. The non-trivial topology simply implies that there is no available scattering channel and the edge conductance is predicted to be exactly $e^2/h$. If time-reversal symmetry is broken, it is no longer guaranteed that gapless edge states persist at the edge. Even if they do, the conductance may deviate from the quantized value due to impurity scattering. This has been observed in WTe$_2$ where the presence of an external magnetic field significantly reduces the edge conductance\cite{Wu2018}. Likewise, the presence of magnetic impurities at edges which may lead to broken time-reversal symmetry\cite{PhysRevLett.102.156603, PhysRevB.95.115430, PhysRevB.97.235402} and destroy the topological edge states. Finally, time-reversal symmetry may be broken \emph{spontaneously} by the presence of magnetism without introducing any impurities. This possibility seems to be largely overlooked in the literature although it by no means is an unlikely scenario. For example, first principles calculations have shown that edges of a monolayer MoS$_2$ in the 2H phase acquires magnetic edge states although the bulk 2D material is non-magnetic\cite{PhysRevB.80.125416}. If time-reversal symmetry is spontaneously broken at an edge of a quantum spin Hall insulator, the presence of gapless edge states is no longer guaranteed. Even if they persist, the observable consequences of the quantum Hall state could in principle be removed by a suitable edge modification. In the present work, we have used first principles calculations to demonstrate that certain edges of transition metal dichalcogenides may acquire magnetic moments leading to spontaneously broken time-reversal symmetry. The spontaneously symmetry breaking at edges is therefore not a mere academic possibility but could destroy the gapless edge states that are typically assumed to be protected by topology.

It is highly desirable to be able to change the topological index of insulators by external means. This would imply that gapless surface states can be removed or introduced at a given edge which may form the basis of one-dimensional transistors. In the case of transition metal dichalcogenides, it has been demonstrated that an external electric field can induce a transition from the quantum spin Hall state to a trivial state\cite{Qian2014}. External electric fields may thus comprise a means to switch between conductive and insulating edges. However, the calculations were based on a tight binding model and constitutes a proof of concept rather than an actual quantitative prediction. In the present work, we have performed full first principles calculations of transition metal dichalcogenides in electric fields. We verify that electric fields can be used to switch the topological state in these materials but find that the magnetic states remain gapless.

The paper is organized as follows. In the second section, we introduce the calculational method and in the third section, we introduce the materials and types of edges that are investigated. The fourth section contains the results on the MoS$_2$ edges and the fifth section contains the results on three other TMDs. In the final section, we investigate the effect of gating on the three MoS$_2$ edges.

\section{\label{sec:methos}Method}

\begin{figure}
	\centering
    \includegraphics[width=1\columnwidth]{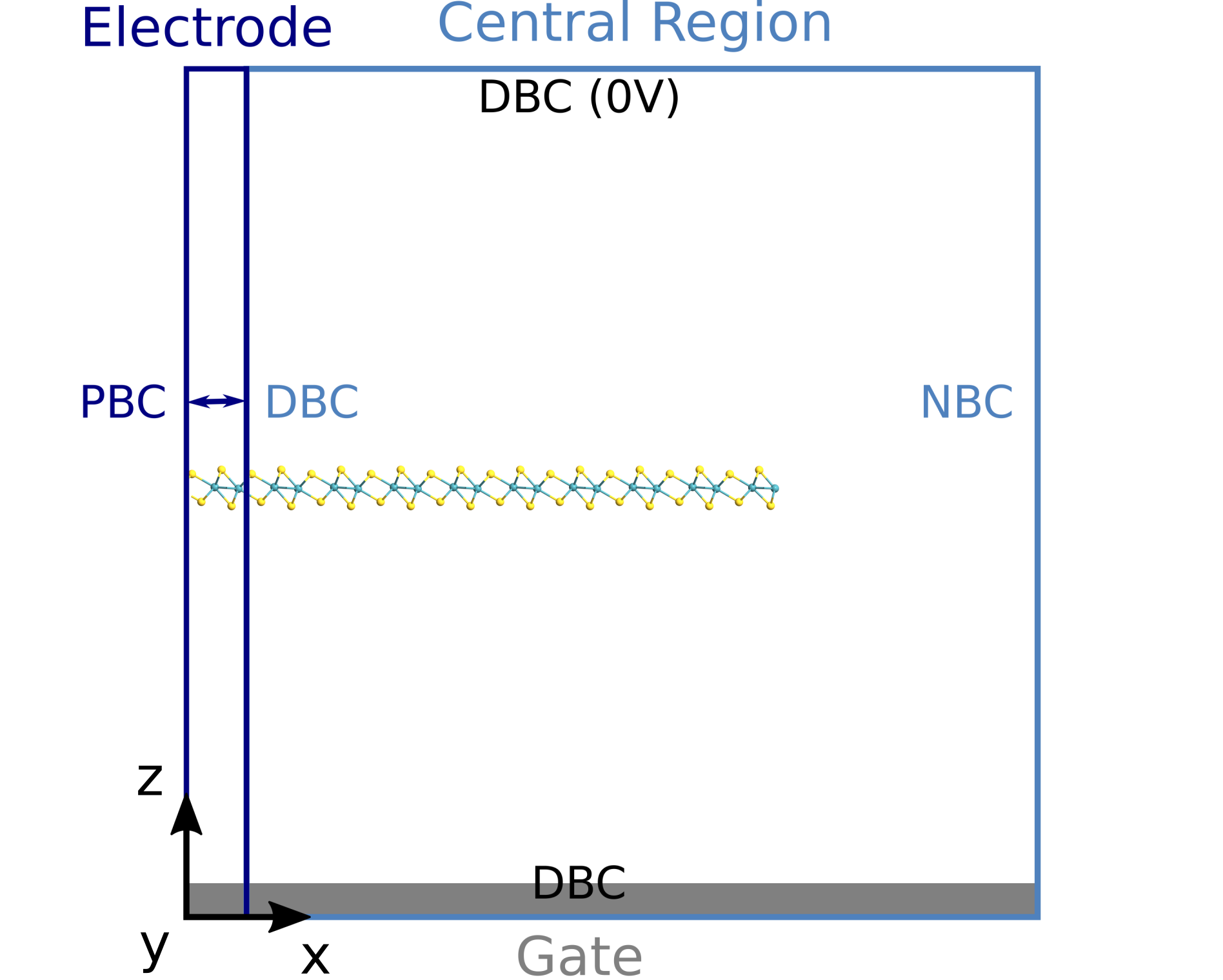}
    \caption{\label{fig:config} The \textit{SurfaceConfiguration} of a ML 1T' MoS$_2$ on top of a gate along with the applied boundary conditions. PBC, DBC, and NBC refers to different boundary conditions as explained in the text.}
\end{figure}

\begin{figure*}
	\centering
    \includegraphics[width=2\columnwidth]{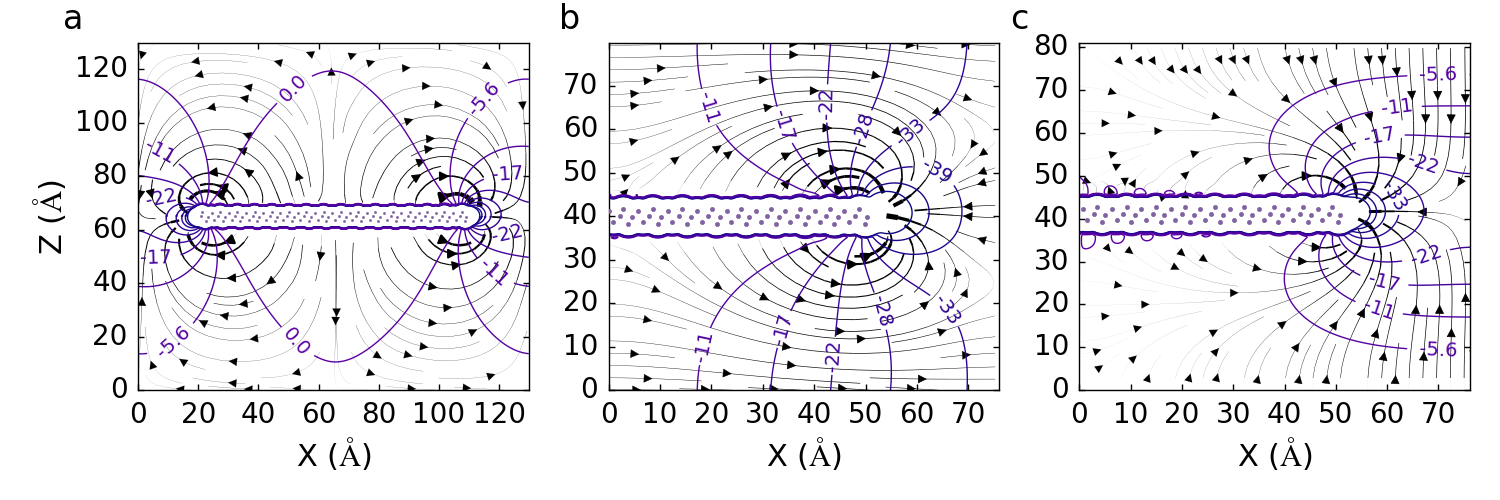}
    \caption{\label{fig:field} The equipotential- and field lines of the effective potential of \textbf{a} a nanoribbon calculation, \textbf{b} a surface configuration without a metallic gate, and \textbf{c} the surface configuration shown in Fig \ref{fig:config} with a grounded metallic gate. The equipotential lines are given in meV.}
\end{figure*}

The calculations are carried out using DFT\cite{DFT1,DFT2} and the Surface Green's Function method as implemented in QuantumATK\cite{surfconfig}. We consider three different configurations in this study (Fig \ref{fig:field}): \textbf{a}, a nanoribbon configuration with two edges and periodic boundary conditions, \textbf{b}, a novel kind of surface configuration (\textit{SurfaceConfiguration}\cite{surfconfig}) with a single edge and a semi-infinite electrode, and \textbf{c}, the surface configuration with a gate below. The nanoribbon configuration is used to compare to the results of a surface configuration in order to pin-point the effect of having a single edge instead of two edges. The gated configuration is used to investigate the response of the edge to an electric field. 

To describe the single edge of the monolayer (ML), we use a surface configuration as shown in Fig. \ref{fig:config}. This type of configuration consists of two regions; the electrode and the central region. The electrode is used to describe the bulk properties of the system and the central region contains all the information on the physics of the edge. The calculation is done in two steps, first a DFT calculation is performed on the isolated and periodic electrode and then a DFT calculation using the Green's function method is performed for the central region. The second calculation uses Green’s functions to couple the central region to the electron reservoir of a monolayer which is periodic in the x- and y-direction (2D crystal). The electron reservoir is included through the self-energy matrix of the semi-infinite electrode. This self-energy matrix is created from the Hamiltonian of the 2D crystal by a recursive method. More details on the Surface Green’s function method can be found in Ref.\cite{surfconfig}. The length of the ML in the central region is determined by converging the Hartree potential wrt. length and making sure that it matches the periodic potential of the electrode. A ML length of 50 \AA\,is adequate for all the systems investigated here.

Special care needs to be taken when describing the electrostatics of 2D interfaces or surfaces. The equipotential- and corresponding field lines of the effective potential of the three configurations are shown on Fig \ref{fig:field}. This illustrates the field which is created by the 1D dipole (monopole) line along the nanoribbon (single edge). Such a field decreases logarithmically and therefore reaches far out into the vacuum region of the system. 

Different treatments are needed for the three types of configurations. In the nanoribbon calculation, we have periodic boundary conditions (PBCs) for all directions. This means that a sufficient amount of vacuum must be applied both above and below the ML and to the left and right of the ribbon in order to avoid the spurious interaction between the periodic images of the ribbon. In the case of the surface configuration without a gate, where we only include a single edge, there is also the issue that there is a slope in the potential above and below the ML. This can be seen on Fig \ref{fig:field}b where the equipotential lines are perpendicular to the top and bottom of the cell. The slope represents the two distinct electronic levels to the right and to the left of the cell. To the left, the work required for an electron to leave the 2D crystal will be determined by the work function of the crystal. To the right, the work required for an electron to leave the edge will be influenced by the created monopole at the edge and therefore not be equal to the work function of the 2D crystal. The effect is that a macroscopic field is set up which persists even infinitely far away. This is an inherent issue of 2D systems\cite{2dint} whereas a 3D system with a 2D mono- or dipole only experiences a local potential shift at the surface or interface. To correctly describe the potential slope and to avoid that it affects the electronic levels, we require that the cell is as high in the $z$-direction as it is long in the $x$-direction. This ensures that the potential has an approximately linear slope at the top and bottom at the cell.

In the case of the gated system, an unconventional set of boundary conditions is chosen for solving the Poisson equation. This is done to accommodate the gate construction and to avoid interactions between images. We apply a Dirichlet boundary condition (DBC) at the top and bottom of the cell so that the potential is fixed in the gate and grounded at the top of the cell. This is analogous to placing the surface between the two plates of a parallel-plate capacitor. For the electrode calculation, we use standard PBCs in all other directions. For the central region, on the other hand, the potential at the left side of the cell is fixed to the value of the potential of the electrode using a DBC and the potential at right side is allowed to converge to the potential in the vacuum region by applying a Neumann BC (NBC). All of these boundary conditions are summarized in Table \ref{tab}.

\begin{table}
\centering
\begin{ruledtabular}
\begin{tabular}{ccc}
 & Electrode & Central region \\   \cline{1-3} \vspace*{-1mm} \\ \vspace{1mm}
$-x$ & PBC   & DBC   \\ \vspace{1mm}
$+x$ & PBC   & NBC   \\ \vspace{1mm}
$-y$ & PBC   & PBC   \\ \vspace{1mm}
$+y$ & PBC   & PBC   \\ \vspace{1mm}
$-z$ & DBC   & DBC   \\ \vspace{1mm}
$+z$ & DBC   & DBC   \\ 
\end{tabular}
\end{ruledtabular}
\caption{\label{tab}Boundary conditions used to solve the Poisson equation for the gated surface configuration.}
\end{table}

Applying DBCs both at the top and bottom of the cell, forces the potential to be zero at those boundaries. The equipotential lines are correspondingly pushed away from these borders and the potential shift is present along the right border of the cell as it can be seen on Fig \ref{fig:field}c. Our investigations show that, if the cell keeps the same height as the ungated system and has enough vacuum to the right s.t. the potential has converged to a constant value when going from left to right, then the electronic levels of the system are equivalent to those of the ungated system.

We apply the Perdew-Burke-Ernzerhof (PBE)\cite{PBE} xc-functional including SOC and non-collinear spin. The wave functions are expanded as a linear combination of atomic orbitals using SG15 pseudopotentials\cite{SG15} together with the SG15-Medium basis set\cite{surfconfig}. The occupations are described using a Fermi-Dirac occupation function with an electronic temperature of 300 K. All the structure relaxations are done without SOC and to a force tolerance of 0.02 eV/\AA.

\section{Materials}

\begin{figure}
	\centering
    \includegraphics[width=1\columnwidth]{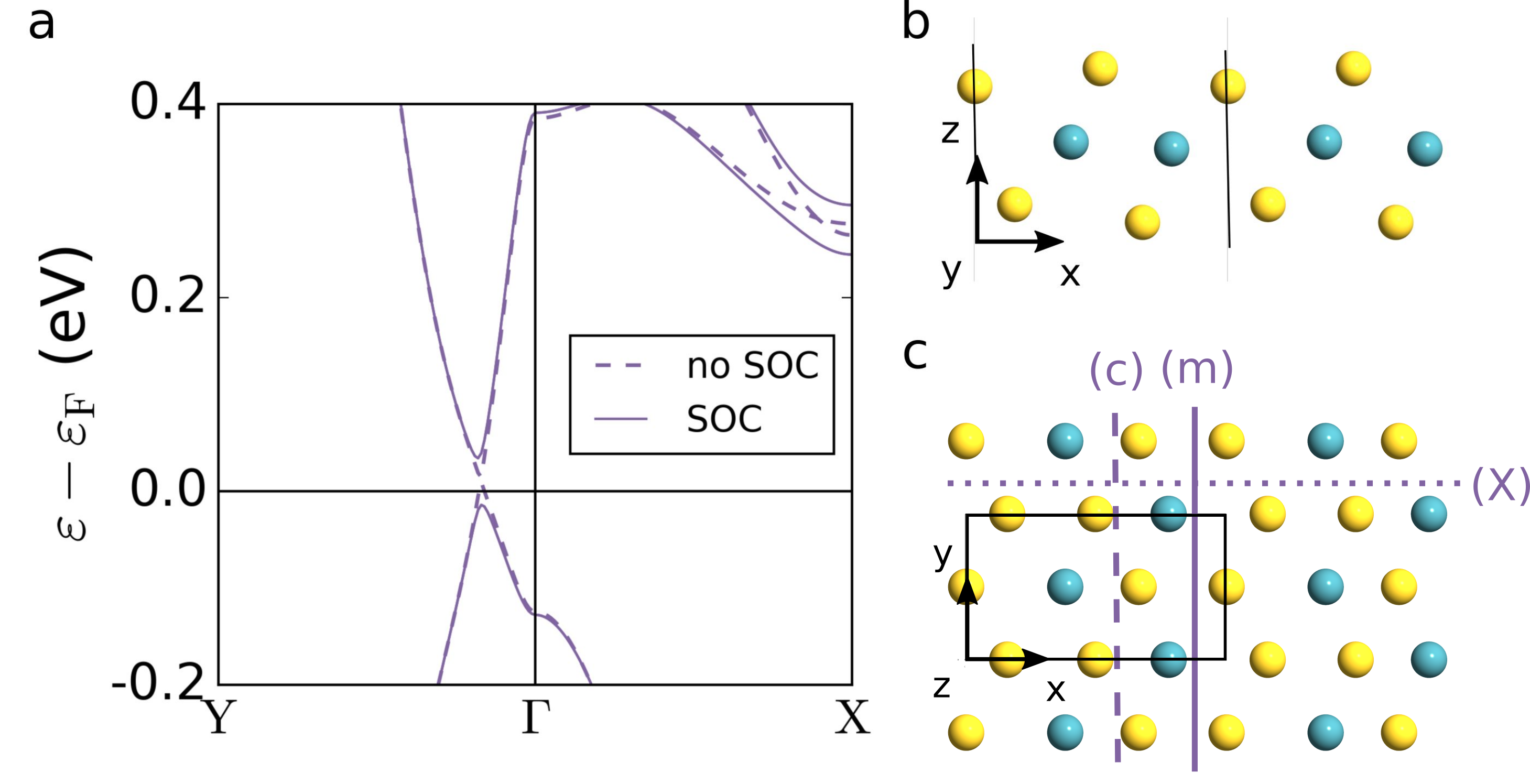}    
    \caption{\label{fig:1} \textbf{a} The band structure of ML MoS$_2$ in the 1T' phase with and without spin-orbit coupling, \textbf{b} Side view of the 1T' phase, \textbf{c} Top view of the 1T' phase showing the three investigated edges.}
\end{figure}

Of all the 2D TMDs, MoS$_2$ is by far the most studied material. This is largely due to the fact that it is easily available in the form of the mineral molybdenite. We will therefore begin by investigating the edges of a MoS$_2$ ML in the 1T' phase. The 1T' phase possesses a small band gap of 80 meV\cite{1TMoS2} due to spin-orbit splitting of the bands occurring between the $\Gamma$ and Y point in the BZ. On Fig \ref{fig:1}a, the band structure is shown with and without spin-orbit coupling to illustrate this. Note, that we obtain a gap of 48 meV which is only slightly lower than the experimental value obtained using STS\cite{1TMoS2}. We will consider 3 different edges, as indicated on Fig \ref{fig:1}c. The (X) edge is a cut along the $x$-direction and the (m) edge and (c) edges are cuts along the $y$-direction. The $y$-cut can be made in several different ways, but these two edges represents the must stable ones for the Mo and single-S terminated kinds respectively, see the supplementary material\cite{supl} for a stability analysis. The k-point sampling of the (m) and (c) edge calculations are: 1 x 11 x 1 k-points for the nanoribbons, 11 x 1 for the surface configurations, which are non-periodic in the $x$-direction, and 401 x 11 x 1 for the periodic 2D crystals. The k-point sampling of the (X) edge is 6 x 1 x 1 k-points for the nanoribbon, 6 x 1 for the surface configuration, which is non-periodic in the $y$-direction, and 6 x 401 x 1 for the periodic 2D crystal.

Even though MoS$_2$ is the most studied TMD, it is not the most studied in the 1T' phase since this is unstable wrt. the conventional 1H phase of the ML. Studies of the 1T' ML TMD phase have primarily been based on WSe$_2$\cite{WSe2,WSe2Nat} and WTe$_2$\cite{Fei2017, Tang2017, Wu2018} where the 1T' phase is more stable. We have therefore included these two materials in our investigations. Finally, we include MoTe$_2$ which has been studied as a material for field-effect transistors where few layered 1T' and 2H phase materials are used as the contact and channel material, respectively\cite{Sung2017}.

\section{\label{sec:MoS2results} The edges of 1T' MoS$_2$}

\begin{figure}
	\centering
    \includegraphics[width=1\columnwidth]{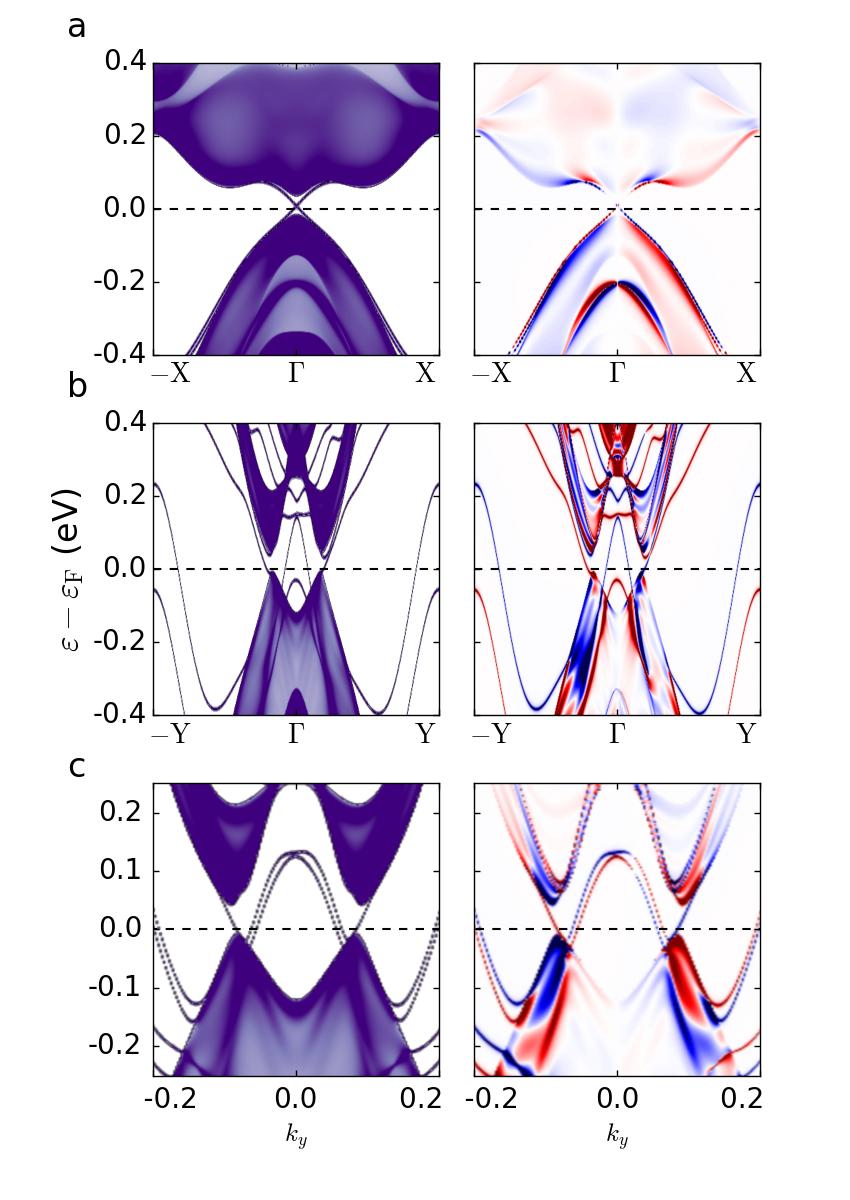}    
    \caption{\label{fig:dos} The electronic bands for \textbf{a} the (X) edge, \textbf{b} the (m) edge, and \textbf{c} the (c) edge of monolayer 1T' MoS$_2$. \textbf{left} side shows the sum of all spin components and \textbf{right} side shows the spin-polarized density of states wrt. the y-direction (\textbf{a}) and the x-direction (\textbf{b} and \textbf{c}).}
\end{figure}

We begin by studying the electronic bands of the three different edges of MoS$_2$. This is done by calculating the density of states for the k-points along the $Y \to \Gamma \to Y$ path of the Brillouin zone. The band structures can be seen in Fig \ref{fig:dos} both for the total number of states and  for the spin-polarized states weighted by the spin component in the x- and y-direction respectively. The solid areas represents the bulk states and the edge states can be identified as the isolated and spin-polarized bands. All three edges host gapless edge states. The (X) edge has 2 edge states which are degenerate two-by-two in the high symmetry points of the Brillouin zone. This is exactly what we would expect for the topologically protected edge states of a topological insulator. However, the behavior is different for the other two edges, for which time-reversal symmetry is broken. The effect is most pronounced for the (m) edge and can be seen on Fig \ref{fig:dos}b where the spin polarization is equal instead of opposite for any $k$ and $-k$ pair of the edge state. Furthermore, it can be seen that the degeneracy at the high symmetry points is lifted and the bands are separated by approximately 0.25 eV at the $Y$ and $-Y$ points. For the (c) edge, the effect is more subtle and can only be seen for the edge states close to the Gamma point between the two dips of the conduction band. Here, the spin polarization changes, not at the $\Gamma$ point, but at a small positive value of $k$ going in the $Y$ direction. 

\begin{figure}
	\centering
    \includegraphics[width=0.95\columnwidth]{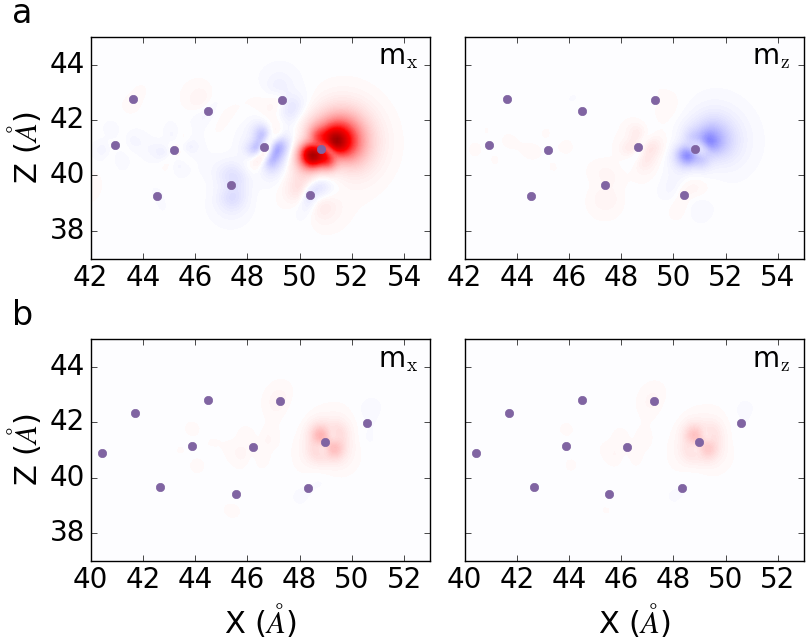}
    \caption{\label{fig:mag} The magnetization density wrt. the $x$ and $z$ axis of \textbf{a} the (m) edge and \textbf{b} the (c) edge of MoS$_2$.}
\end{figure}

The origin of this effect is, that the time-reversal symmetry is spontaneously broken as a magnetic state is formed at the edge. This can be illustrated by a plot of the magnetization density as shown in Fig \ref{fig:mag}. Note, that the spin configuration is non-collinear and we therefore have a magnetization density wrt. each spatial direction. The figure only shows the $x$ and $z$ components, since the $y$ component is zero. The magnetic moments point to the right and downwards (upwards) for the m (c) edge and have most weight on the last Mo-atom of the surface. By investigating the projected density of states near the $\Gamma$ point, we find that the edge states primarily stems from the Mo d-orbitals and S p-orbitals with the approximate relative weight of 3:1 for the (m) edge and 5:2 for the (c) edge.

From the nanoribbon calculation with two (m) type edges (as illustrated on Fig \ref{fig:field}a), it can be seen that the magnetic state points in the same spatial direction on both edges. This is also in contrast to the QSHI where two spin channels going in opposite directions resides on each edge. The electronic bands of the nanoribbon are symmetric wrt. k since the two edges are symmetric so that the edge state at each $-k$ point from the left edge has an identical state at the $k$ point on the right edge. A spin resolved DOS will off course still reveal the breaking of the time-reversal symmetry. Nevertheless, this underlines the strength of the surface configuration for investigating single edges one at a time. More results from the nanoribbon calculation can be found in the supplementary material\cite{supl}. 

In conclusion, we find that the time-reversal symmetry can be broken on the edges of 1T' monolayer MoS$_2$ and that it leads to magnetic states which behaves very differently from the expected topological states. This means that the gapless states on these edges are not topologically protected and one could imagine that they could be removed by reconstructions of the edge.

\section{\label{sec:Wresults} Magnetic edges on 1T' TMDs}

We now turn our attention to the other three TMDs of the study; WSe$_2$, WTe$_2$, and MoTe$_2$. In the investigations of these materials, we use the configuration shown on Fig \ref{fig:field}b since we do not wish to apply a field to these edges. The resulting electronic bands can be seen in the supplementary material<cite{supl}. All investigated edges exhibit edge states. To quantify the magnetism, we calculate the magnetization vector,

\begin{equation}
    \vec{m} = \frac{1}{N_M} \int{}{}{\vec{m}(\vec{r}) \, \mathrm{dV}}, 
\end{equation}

and the total magnetization,

\begin{equation}
    m = \frac{1}{N_M} \int{}{}{\sqrt{\vec{m}^2(\vec{r})} \, \mathrm{dV}}.
\end{equation}

 per transition metal atom on the edge,  $N_M$. Note that for all three investigated edges, $N_M=1$. The results for the (m) and (c) edges of all the 4 investigated materials are seen in table \ref{tab:mag} along with the distance between the valance and conduction bands. We note that MoTe$_2$ and WTe$_2$ are metallic since the valence and conduction bands overlap in energy. However, the bands are separated in k-space which means that the character of the electronic structure is similar to the other materials and that a topological index can be defined. The magnetic edge state is present for MoS$_2$, MoTe$_2$, and WTe$_2$ but missing for WSe$_2$.
As seen in the table, the magnetic (m) edges show a higher magnetism than the magnetic (c) edges for the same material. This can be understood through the different stoichiometry of the two termination. At the (m) type edge, the outermost metal atoms are completely exposed and missing 3 of the nearest chalcogenide atoms. This means that the metal atoms are hindered in transferring electrons and end up with a small excess of electrons compared to a metal atom in the 2D crystal. These excess electrons will be filled into the d-band resulting in a higher DOS at the Fermi level. This higher DOS results in magnetism in accordance with the Stoner criterion.  At the (c) type edges, the outermost metal atoms only lacks a single chalcogenide neighbor. The bonds are therefore more saturated, the d-band less filled, and the magnetism smaller. From the projected density of states, we find that the edges which become magnetic have edge states where the d-orbitals have a weight which is larger than twice the weight on the p-orbitals. This agrees with the explanation above and shows that the magnetism arise when d-type states dominate the Fermi level.

\begin{table}
\centering
\begin{ruledtabular}
\begin{tabular}{cccccc}
         & \multirow{2}{*}{$E_g$\footnote{Difference between valance band maxima and conduction band minima}}  & \multicolumn{2}{c}{(m) edge} & \multicolumn{2}{l}{\hspace{3mm}(c) edge} \\ 
         &   &  \multicolumn{1}{c}{m} & \multicolumn{1}{c}{$\vec{m}$} & m & $\vec{m}$\\
         \cline{5-6} \cline{3-4} \cline{5-6} \vspace*{-1mm} \\ \vspace{1mm}
MoS$_2$  & 48 meV & 6.7 & (0.26,0,-0.069) &  0.82 & (0.035, 0, 0.028)   \\ \vspace{1mm}
MoTe$_2$ & 0 meV  & 17  & (0.75, 0, 0.26)   & -     &                    \\ \vspace{1mm}
WSe$_2$  & 10 meV & -    &                  & -     &                    \\ \vspace{1mm}
WTe$_2$  & 0 meV  & 5.7 & (0.23, 0, 0.10)  &  1.0 & (0.035, 0, -0.030)  \\ 
\end{tabular}
\end{ruledtabular}
\caption{\label{tab:mag} Calculated band gaps, total magnetization in $\mu_B \times 10^{-2}$ and magnetization vector in $\mu_B$ pr. transition metal atom on the edge for the 4 different 1T' monolayer TMDs.}
\end{table}

\section{The effect of gating}

\begin{figure}
	\centering
    \includegraphics[width=1\columnwidth]{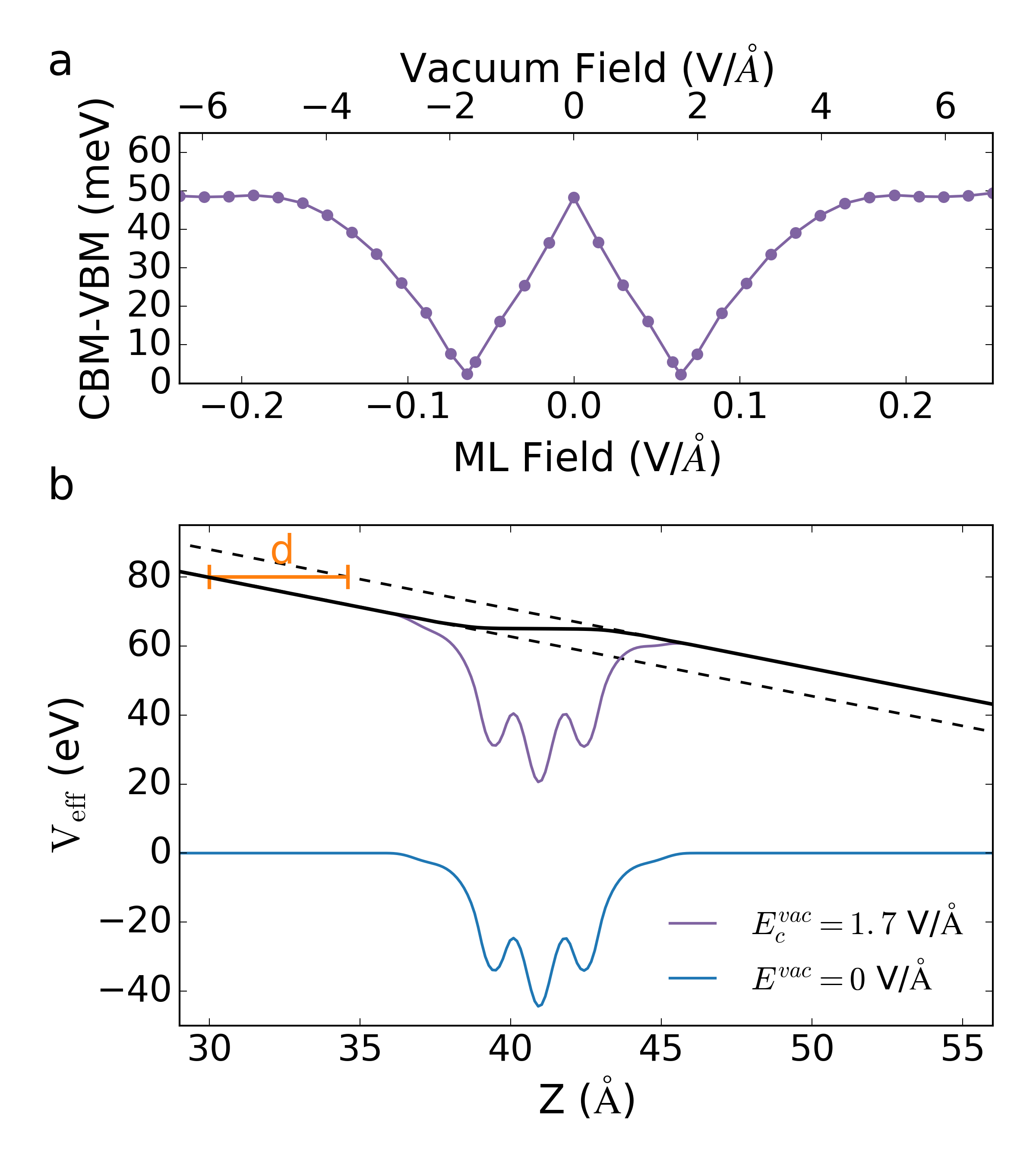}
    \caption{\label{fig:bg_vs_E} \textbf{a} Gap between valence and conduction band of ML 1T' MoS$_2$ as function of the effective potential in the vacuum region and the ML, and \textbf{b} Effective potential of the zero bias and critical field configurations.}
\end{figure}

\begin{figure}
	\centering
    \includegraphics[width=1\columnwidth]{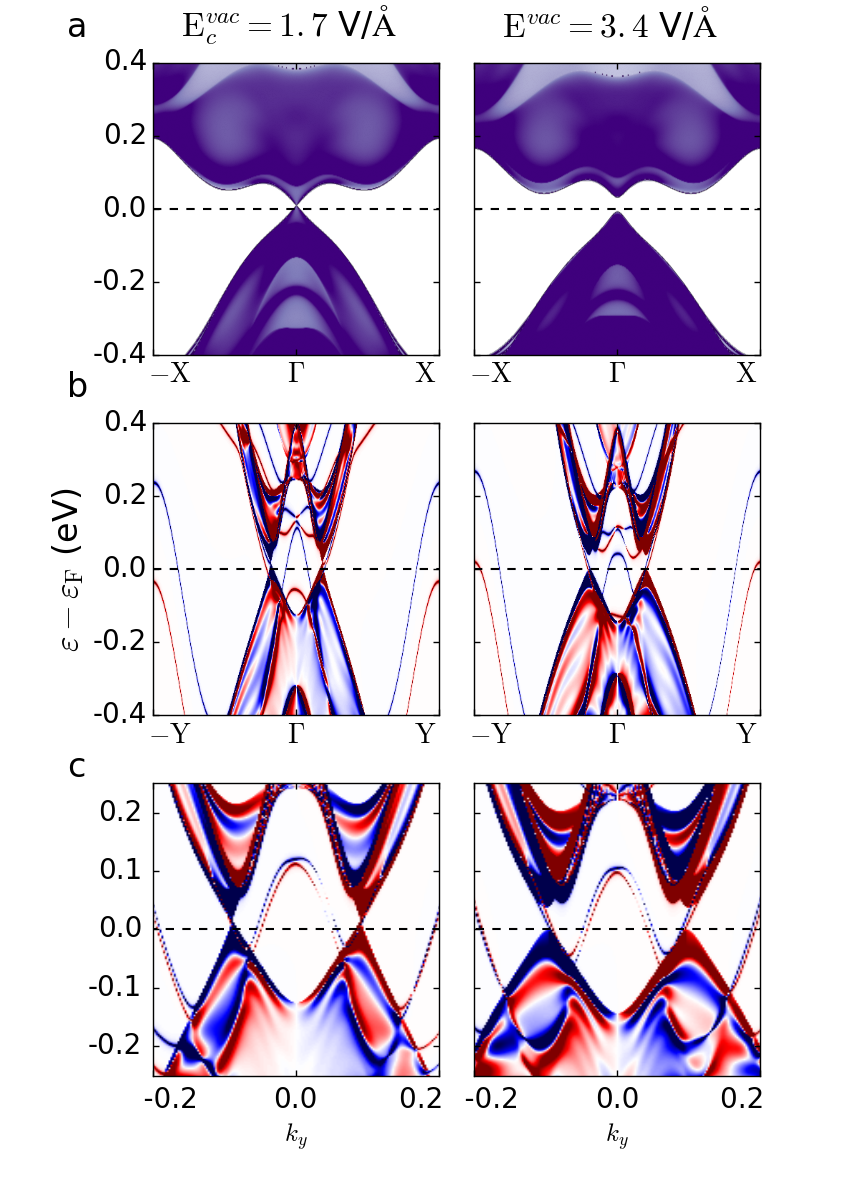}    
    \caption{\label{fig:dos_gate} Response from of the electronic bands of 1T' MoS$_2$ when applying a perpendicular field. \textbf{a} the (X) edge, \textbf{b} the (m) edge, and \textbf{c} the (c) edge.}
\end{figure}

It has previously been shown, based on electronic structure calculations\cite{Qian2014}, that the metallic edge states of the 1T' TMD QSHIs can be switched off by applying an electric field perpendicular to the monolayer. This field closes the band gap and thereby changes the topological state of the material. At a higher field strength, the gap opens again, the system has become topologically trivial, and the gapless edge states are removed. We do similar investigations for the three edges of MoS$_2$ using full self-consistent DFT to describe the response to the electric field. To find the critical field at which the gap closes, calculations are performed for a range of field strengths perpendicular to the electrode cell of the configuration. For these calculations, we use a k-point grid of 6 x 11 x 1. This describes the response of the 2D crystal and can be seen on Fig. \ref{fig:bg_vs_E}a. The field is applied by adding an external potential as a shift between the gate and the top of the cell and then running a self-consistent DFT calculation. The corresponding field strength in the vacuum region and in the monolayer can be found as the slope of the effective potential in these regions. The zero field potential is subtracted in order to extract the slope in the ML region. The effective potentials for zero bias and the critical potential of 130 V is shown on Fig \ref{fig:bg_vs_E}b along with the difference between the two. The resulting critical fields are $E_c^{vac}= 1.7$ V/\AA\, in the vacuum and $E_c^{ML}= 6.4 \times 10^{-2}$ V/\AA\, in the ML. The critical field in the ML is about half the size of the reported value from the previous study\cite{Qian2014} where the field is added as a correction to the diagonal elements of the Hamiltonian. The large difference between the vacuum field and and field inside the material shows that the monolayer strongly screens the field. In particular, the longitudindal part of the dielectric constant in the $z$-direction can be estimated as the ratio between the vacuum field and the ML field, $\ve_{ML} = E^{vac}/E^{ML}=27$. 

As it is indicated by the dashed lines on Fig. \ref{fig:bg_vs_E}b, a characteristic distance over which the potential on either side of the monolayer is shifted can be found. Since the induced field must vanish in the vacuum region, it is straightforward to verify that this shift can be written as 
\begin{align}
    d =& -\frac{1}{E^{vac}}\int dz\Big[E^{ML}(z)-E^{vac}\Big]=\frac{4\pi}{E^{vac}}\int dzP_{\perp}(z)\notag\\
    =&\frac{4\pi P_{\perp}^{2D}}{E^{vac}}=4\pi \alpha_{\perp}^{2D},
\end{align}
for any applied field. $P^{2D}_{\perp}$ is the 2D polarization perpendicular to the monolayer i.e. the dipole per unit area. Except for a factor of $4\pi$, we can thus identify $d$ with the perpendicular component of the 2D polarizability $\alpha^{2D}_\perp\equiv P^{2D}_\perp/E_\perp$. For the case of MoS$_2$ in the 1T' phase shown in Fig. \ref{fig:bg_vs_E} we get $d=4.6$ \AA. This compares reasonably well with the value of $\alpha_\perp^{2D}$ calculated using RPA for the Computational 2D Materials Database\cite{c2db}, which yields $d = 5.40$ \AA.

Having determined the critical field, we perform calculations on the three edges with the critical field and twice the critical field applied. The electronic bands can be seen on Fig \ref{fig:dos_gate}. The edge which has been investigated previously in Ref.~\onlinecite{Qian2014} is the non-magnetic (X) edge. For this edge, we see the expected behavior. Above the critical field, the gapless states have disappeared and no conducting channels are available. For the two magnetic edges, on the other hand, the metallic edge states persist relatively unaltered above the critical field. For the case of the magnetic edges, it is therefore not possible to remove the conducting states by applying a field. Due to the weak coupling between the magnetic and electronic degrees of freedom, the total magnetization also remains relatively unaltered with respect to the field strength.

\section{\label{sec:conclusion}Conclusion}

Using DFT and the NEGF method, we have studied the single isolated edges of four different monolayer TMDs in the 1T' phase. We find that several of the edges show breaking of the time-reversal symmetry and exhibits magnetic edge states. This means that the gapless edge states of these edges are no longer protected against impurity scattering and that they in principle could be removed by edge modifications. The total magnetization varies over the four materials and is strongest for one of the MoTe$_2$ edges with a value of 0.17 $\mu_B$ pr. unit cell.
We have also studied the response of the edge states of MoS$_2$ when applying a perpendicular field. Self-consistent DFT calculations show that only the topologically protected gapless edge states are removed above a critical field while the gapless magnetic edge states remain relatively unaltered.

\vspace*{0.1cm}

\begin{acknowledgments}
This work is partly funded by the Innovation Fund Denmark (IFD) under File No. 5189-00082B. The authors acknowledge Julian Schneider and Troels Markussen for technical help with the calculations and interpretation of the results.
\end{acknowledgments}

\bibliography{refs}

\end{document}